\documentclass[darkmode,twocolumn]{aastex631dm}

\newcommand\aastex{AAS\TeX}

\usepackage[caption=false]{subfig}
\usepackage{booktabs}

\submitjournal{all journals it may concern}

\shorttitle{The Case for Dark Mode in Journals}
\shortauthors{Corcoran \& Corcoran}

\begin{document}

\title{``My Rhodopsin!'': Why Adding Dark Mode to Journals Could Make Us All Better Astronomers}

\author[0000-0002-2764-7248]{Kyle A. Corcoran}
\affiliation{University of Virginia, Department of Astronomy, 530 McCormick Rd., Charlottesville, VA 22904, USA}

\author{Ellorie M. Corcoran}



\begin{abstract}
The digital age has sparked a revival in the use of ``dark mode'' (DM) design in many everyday applications as well as text editors and integrated developer environments.  We present the case for adding a DM theme to astronomical journals, including a modified class file that generates the theme you see here as a potential option.  DM themes have many beneficial attributes to a user such as saving battery power and reducing screen burn-in on devices with OLED screens, increasing figure hopping efficiency, pairing well with colorblind-friendly palettes, and limiting rhodopsin loss while observing.  We analyzed iPoster design trends from AAS 237 and 238 to gauge the possible reception of our DM theme, and we estimate that at least 35\%, but likely closer to 42\%, of the community would welcome this addition to journals.  There are some drawbacks to using a DM theme when reading papers, including increased ink usage when reading in a print medium and some diminished legibility and comprehension in low-light conditions.  While these issues are not negligible, we believe they can be mitigated, especially with a paired submission of both a DM and traditional, ``light mode'' manuscript. It is also likely that many of us will become better astronomers as a result of adding DM to journals.
\end{abstract}

\keywords{User Interface Design: Dark Mode --- Accessibility in Science --- Humor in Science}


\section{Introduction} \label{sec:intro}

Negative polarity (dubbed ``dark mode'') displays have seen a rise to prominence again in recent years on the heels of the introduction of such user interfaces (UIs) by various companies such as Apple.  The concept of light text on a dark background is not novel, though.  Early iterations of computers used cathode ray tube monitors that could only display lighter (usually green) text on a dark background.  Advancement in display technology eventually gave rise to the concept of ``what you see is what you get'' (WYSIWYG) in the 1980s, and many application and operating system UIs adopted ``light mode'' (LM) themes that mimic printed-paper style as a result.  The WYSIWYG movement obviosuly did not erase dark mode (DM) themed interfaces; however, the standard for UIs did firmly side with LM for many years.

In our increasingly digital world, many applications\footnote{\url{https://darkmodelist.com/} by \citet{Azimov2020} provides a list\\of 300 examples} that we commonly interact with, both scientific and not, now have options for or are created using DM design.  Text editors and integrated developer environments such as Atom, PyCharm, Visual Studio, Sublime Text, etc. are good examples of this as they all default to UIs with DM themes.  Computer terminals are often in DM, too, mimicking the original display style pre-WYSIWYG.  This is true for many flavors of Linux, true in the case of macOS when the system UI is set to DM, and even Windows got something right in a way -- albeit color theory would probably advise picking any dark-background color other than pure black \citep{Hoober2020} -- with the command prompt.  Social media apps such as Twitter and Instagram are prominent purveyors of DM UIs, although they often have both DM and LM options and leave the choice to the user based on preference.

Unlike the rapid renaissance of DM design, the literature has been slow to meet consumer needs for validation of their UI preferences.  Searching on the term DM in Google Scholar only yields two relevant results on the first page as an example.  The same exercise performed on the Astrophysics Data System (ADS; in DM, of course) and arXiv is not as successful.  While this leaves mostly only anecdotal reasons for inverting one's opinion in addition to their UI, they are still worth considering in the context of the literature that does exist.  Mostly anecdotal or not, a DM theme and the rationale behind certain elements of the design could make all of us better astronomers in the long run.

Here we present the case for bringing a DM theme to scientific journals, especially arXiv pre-prints and those under the American Astronomical Society (AAS) Journal umbrella.  In \textsection\ref{sec:aastexdm}, we showcase an option to create articles in said theme, which the reader has probably noticed by now.  We then outline the potential benefits of displaying articles in this theme in \textsection\ref{sec:benefits}.  We attempt to predict the impact of our theme in the community based on the use of DM and LM themes in astronomy in \textsection\ref{sec:aasposters} using iPosters from previous AAS meetings.  In \textsection\ref{sec:drawbacks}, we comment on potential drawbacks caused by a DM alternative and explain how to mitigate those issues.  Finally, we summarize in \textsection\ref{sec:summary}.

\section{\aastex\ --- darkmode} \label{sec:aastexdm}
To aid in the preparation of manuscripts using a DM theme, we modified the class file of the most recent version of \aastex\ (6.3.1) to introduce the necessary design changes to the document when compiling.  As we still wanted the user to be able to draft documents in the default LM style should they wish to, our DM theme can simply be called by adding the keyword \texttt{darkmode} in addition to your preferred document-style keyword to the \texttt{\string\documentclass} command.  This process is similar to calling \texttt{linenumbers} for drafts submitted to AAS Journals for the refereeing process.  Below we highlight some of the key characteristics of our new theme, and the class file for calling it, \href{run:./aastex631dm.cls}{aastex631dm.cls}\footnote{\url{https://github.com/kacorcoran/aastex-darkmode/}}, can always be further modified to a user's specifications if needed.

\subsection{Theme Colors} \label{ssec:theme}
To construct our proposed DM theme, we first performed quick visual inspections of the themes of several text editors and applications currently on our local machines.  We then sampled the background and text color within those themes using the GNU Image Manipulation Program \citep[GIMP;][]{gimp} eyedropper tool.  Each hex color code was recorded and stored for use later, resulting in a total of six potential background colors and four potential text colors to choose from.  We then used Python to generate a plot containing sample text for each background--text, color combination with the stipulation that the parent themes they were derived from could not be replicated.  As there were three parent-theme pairs in our sample, we were left with 21 possible combinations of background and text colors for a new theme.  

After viewing all options, we decided that we did not like any of them.  We instead chose the one that we found most appealing and again used GIMP to slightly modify the RGB values of both colors ($\sim$3--9 points each on the GIMP scale of 1--100), which resulted in the beautiful theme you see here.  Our adopted background color, which we have dubbed ``\texttt{almost\_atom}'' (\#2D2F34), derives from the ``One Dark'' theme in the text editor, Atom.  The defining characteristic of \texttt{almost\_atom} is that we have slightly warmed the gray up to equilibrium because it was a touch too blue for our purposes herein.  We use the color we have dubbed ``\texttt{brighter\_discord}'' (\#F3F4F3) for the main text color.  Although it is fairly self-explanatory, \texttt{brighter\_discord} derives from the main text color used on the communication platform, Discord, where we have shifted the hue and brightness toward white slightly to provide more contrast between the text and the background.

\subsection{Color Palette for Links} \label{ssec:links}
In addition to defining color macros for the background and text, \href{run:./aastex631dm.cls}{aastex631dm.cls} also contains new colors for links that are better suited to the custom theme.  We define multiple colors that can be chosen to replace the standard \texttt{xlinkcolor} in \aastex; however, we also showcase the possibility of using a palette of colors within the document to better distinguish between the kinds of links being presented.  Although we found no studies that explore potential benefits/drawbacks of colorblind-friendly (CBF) palettes in DM, we believed many of the CBF palettes could be cohesive with our theme.  We also feel it is important to bring up these topics of accessibility in science when possible, so adopting a CBF palette served double duty in that respect.  \citet{Wong2011} has a fairly extensive CBF palette containing many light hues, and we selected five to represent different link types a user might need when constructing a manuscript.  Our palette is mapped as follows:
\begin{itemize}
    \item \textcolor{cbforange}{CBF Orange (\#E69F00)}: internal links to figures, tables, etc.
    \item \textcolor{cbflightblue}{CBF Light Blue (\#56B4E9)}: bibliography links
    \item \textcolor{cbfyellow}{CBF Yellow (\#F0E442)}: external links such as URLs
    \item \textcolor{cbfgreen}{CBF Green (\#009E73)}: file links
    \item \textcolor{cbfpink}{CBF Pink (\#CC79A7)}: doi links
\end{itemize}
We note that CBF Light Blue would be the natural choice should only one color be used, yet CBF Orange and Yellow would also be adequate choices.  We also note that when preparing this manuscript we modified a reference style from GitHub\footnote{\url{https://github.com/yangcht/AA-bibstyle-with-hyperlink}} that mimics the publisher-version, doi-link citation style.  This new reference style, \href{run:./aa_url_dm.bst}{aa\_url\_dm.bst}\footnote{included in the same repository as \href{run:./aastex631dm.cls}{aastex631dm.cls}}, can be used to get publisher-looking doi links in a different color, but the \texttt{aasjournal} style will still work just fine, too.  To highlight the benefit of the selected colors, we used a visualization tool\footnote{\url{https://davidmathlogic.com/colorblind}} to display how each color is altered under some common types of color blindness (shown in Figure \ref{fig:cbf_palette}).  It is clear that these colors work both with each other as well as against the background and text color described in \textsection\ref{ssec:theme}.

\begin{figure}
    \centering
    \includegraphics[width=\linewidth]{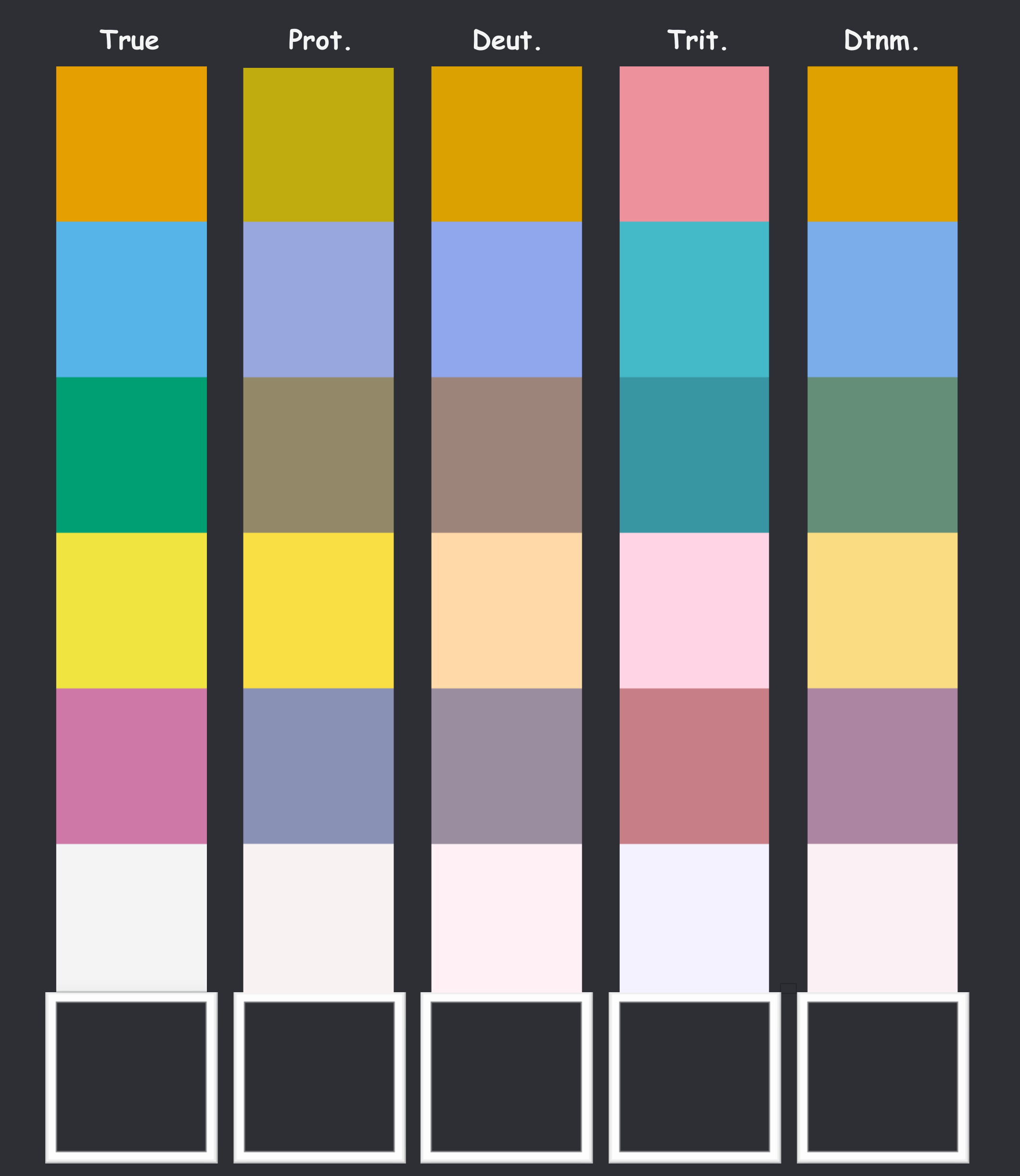}
    \caption{Our selected color palette with each color shown in its true color on the far left as well as how it would appear to someone with a few common types of color blindness.  From left to right, these color blindness or anomaly variants are: protanopia, deuteranopia, tritanopia, and deuteranomaly.  It is worth noting that \texttt{brighter\_discord} does change in color slightly and thus is not purely CBF; however, it still has significant contrast relative to the rest of the CBF palette.}
    \label{fig:cbf_palette}
\end{figure}

\section{Benefits of Using a DM Theme} \label{sec:benefits}
In this section we outline many of the benefits that would come from adding a DM theme to journals.  There are likely more benefits that are not covered here, though, so these should be taken as a pilot study.  These benefits can help to make us better astronomers, too.

\begin{figure}
    \centering
    \includegraphics[width=\linewidth]{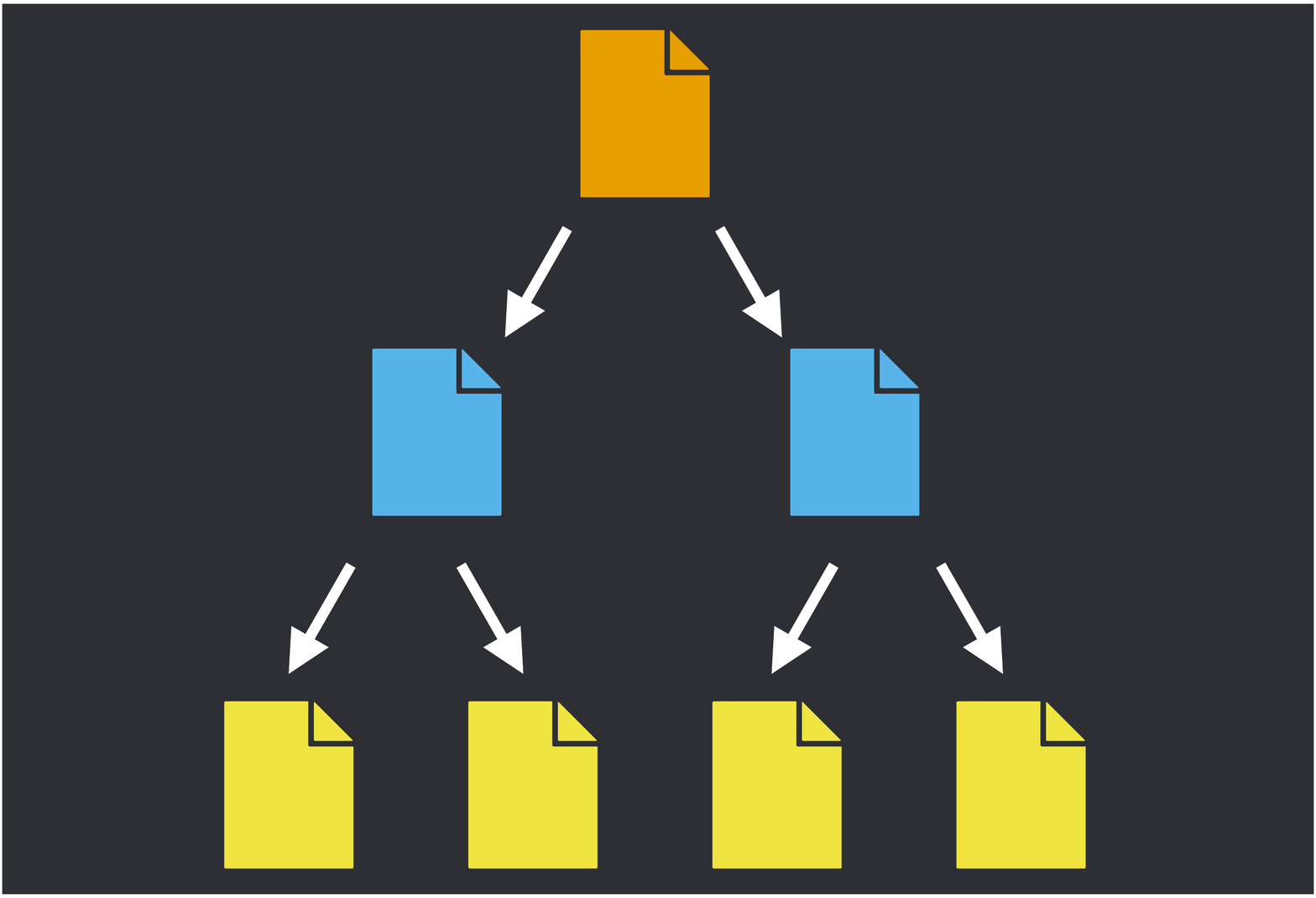}
    \caption{General schematic showcasing how one could discover new papers and topics with the increased battery life provided by a DM theme in journals.  One paper would lead to looking at more papers that in turn would also lead to looking at more papers and so on and so forth.  We estimate users should be able to get about 42\% further in this exercise using a DM theme.}
    \label{fig:pyramid}
\end{figure}
\subsection{Battery Power and Screen Burn-in} \label{ssec:battery}
One of the most well documented benefits of DM themes in the literature is their ability to significantly cut down on battery usage on devices with modern OLED screens \citep[e.g.,][]{Dong2009,Dash2021}.  OLED screens can also suffer from screen burn-in, and DM UIs and themes help reduce this issue during extended periods of use.  These facts can be paired with the benefits of online access to journals and reference managers to bolster an astronomer's reading experience.  Cutting down on power consumption is vital for long sessions of reading papers as it allows one to not be tethered to an outlet to keep your computer alive.  By saving time before needing to plug back in, one can start reading a paper, find a citation to another interesting paper and open it, and find new interesting references to open from those papers, too (see Figure \ref{fig:pyramid} for visualization).  We estimate that this exercise could progress about 42\% more down the branches of the reference tree with the saved power.  Additionally, cutting down the power consumed by having a single instance of an article open means that more articles can then be opened within a browser with the same amount of power (checking that order of operations is left as an exercise for the reader).  Now we need not worry about running our batteries down so quickly keeping all those tabs open with papers that look interesting that we are definitely going to read soon because this time we said we would do it is different.  The reduced screen burn-in also helps to ease worries about accidentally burning in different papers to your display.

\subsection{DM can Reduce Eye Strain} \label{ssec:eyestrain}
Literature surrounding DM themes helping to reduce eye strain in general is sparse and mostly inconclusive; however, \citet{Erickson2020} found that using DM for various reading tasks with a virtual reality (VR) headset led to significantly less visual fatigue.  VR displays generally have better resolution than typical screens, and the headset aspect means the display is very close to the user relative to other displays.  Thus, less eye strain in VR displays may still be a unique case, but in general this particular topic is largely supported by anecdotal claims.  This does make some sense ocularly since the pupil may not have to stay as constricted when looking at DM themes for extended periods of time. 

\subsection{Figure Hopping is More Efficient} \label{ssec:fighopping}
One of the most common tactics we employ to quickly digest papers is ``figure hopping,'' where we may read the abstract, the conclusions, and then, instead of the paper body, we use the figures in the paper to piece together the story of the research being presented.  According to a study by the company, Salesforce, DM themes can help with that method.  They investigated how the theme of an application may affect performance and user preference when viewing charts.  Participants were asked to comment on the design of the UI and certain charts within the app, and the study measured how quickly and accurately the participants responded to questions about the charts.  Looking at charts in DM took less time than the same charts in LM with similar accuracy \citep{fadden_geyer_2018}.  This means that the addition of DM to journals might help readers digest figures even quicker, and they may even save enough time to actually read more of the paper.

\subsection{Plots}\label{ssec:plots}
The following benefits are all specific to preparing plots for a manuscript.  All of these advantages are mostly assuming that \texttt{Matplotlib} \citep{Hunter2007} is being used to create plots; however, the general ideas would translate to other languages or plotting suites.

\subsubsection{CBF Colors Look Fantastic} \label{sssec:cbfcolors}
As we stated in \textsection\ref{ssec:links}, our color palette was chosen because it is both CBF and looks good in general overlaid on our background color, \texttt{almost\_atom}.  Similarly, this palette is particularly useful for plotting.  Many kinds of plots typically only need a few colors to represent the data/model being shown, so keeping the number of colors used smaller is beneficial for the plot being CBF.  The color palette used for links in the document would be a natural choice as it only has five colors that all work cohesively.  Even if no DM theme is ever implemented in journals, this palette can still be used for creating plots without any loss in effectiveness reaching  those with color blindness. 

\subsubsection{A New Suite of Other Possible Colors} \label{sssec:plotcolors}
In addition to the CBF palette, a DM theme more generally opens up the possibility to use many different colors in plots.  In just the named colors of \texttt{Matplotlib}, there are maybe around 30\% of the colors that users can essentially never use.  These colors are simply too light in hue to show up on a white background, too low in contrast to stand out from the white background, or just not as exciting as other colors.  Adding a DM theme to journals allows for the user to get more creative with color choices as well as mixing and matching contrasting colors.  We highlight some of the potential options that one can choose from in Appendix \ref{appendix}. 

\begin{figure}
    \centering
    \includegraphics[width=\linewidth]{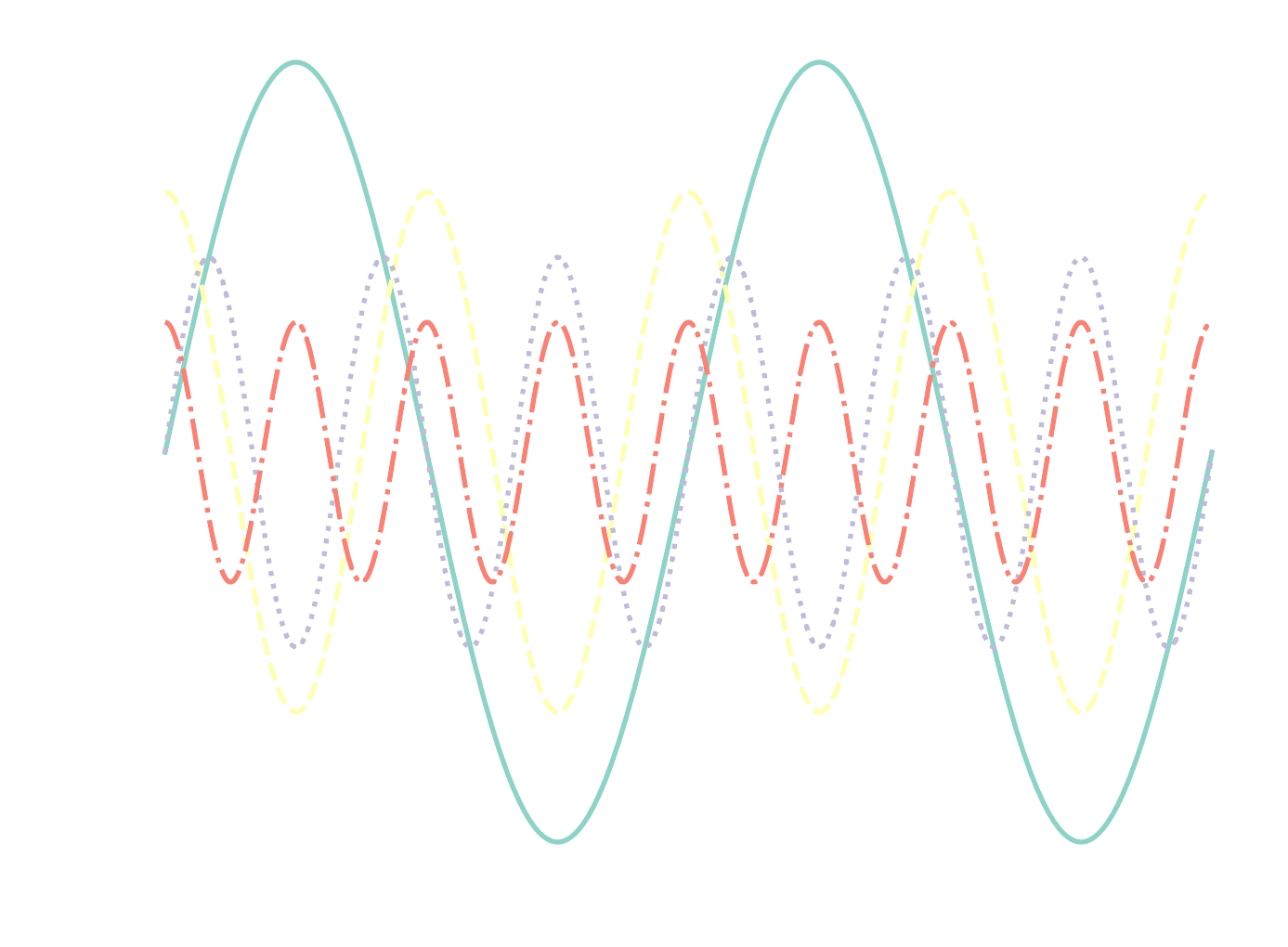}
    \caption{An example of a plot output with the default settings in the \texttt{dark\_background} style in \texttt{Matplotlib}.}
    \label{fig:example}
\end{figure}

\subsubsection{\texttt{Matplotlib} Contains a DM Plotting Style} \label{sssec:darkbgstyle}
This is less an advantage and moreso just pointing out that it would not take signifcantly more effort to change plots to be DM friendly.  One of the style options for the axes of \texttt{Matplotlib} figures is called \texttt{dark\_background}\footnote{\url{https://matplotlib.org/stable/gallery/style\_sheets/dark\_background.html}}, and it plots most elements in white rather than black by default.  To make a plot in this style, the only thing needed is to add the line \texttt{plt.style.use(\textquotesingle dark\_background\textquotesingle)} (assuming one has imported \texttt{matplotlib.pyplot} as \texttt{plt}) to the code above plotting calls.  Then all that is needed to finalize the plot is to add transparency when saving.  An example of a plot constructed using the default parameters of the \texttt{dark\_background} style can be seen in Figure \ref{fig:example}.  If the user does not prepare plots in Python for their manuscript, alternate preparations will need to be made to create the plots.

\subsubsection{Image Formatting} \label{sssec:plotfiletype}
By adding a DM theme to journals, the need for transparent background figures would understandably increase.  While it is true that one could also construct plots with the same facecolor as the journal's background color, that does not get at the point we are trying to make here.  The point is that PNG is an objectively better image format than JPEG even if you take out the fact that it allows for transparency, so adding a DM theme might finally allow journals to accept PNGs.  Similarly, PDFs are likely just the best way to submit figures for both online and print viewing anyway, meaning the faster DM is introduced, the faster more people switch to using PDF as a plotting format.  This one small change could even benefit the papers still prepared in a more traditional LM theme.  We are aware that saving figures in these formats can lead to larger file sizes, but we do not believe this is going to be a substantial issue for most users.  

\subsection{Saving Your Rhodopsin} \label{ssec:rhodopsin}
We have all likely been in a situation where our eyes were well dark adapted -- perhaps asleep and just woke up for example -- and attempted to view a device with a screen that was far brighter than we expected.  This is generally not a pleasant experience.  Even when we are well adjusted to device brightness and in low-light conditions for extended periods of time, we as astronomers need to stay as dark adapted as possible.  Exposure to light depletes our rhodopsin -- a complex protein in the rods of our eyes that aids in low-light sensitivity -- levels \citep{Majewski2021}.  DM themes are far less harsh in outputting light, especially blue light, and may help reduce the amount of rhodopsin lost if reading papers in low-light conditions. This is especially important for astronomical observing as replenishing rhodopsin levels takes around 30 minutes, and there are many other issues that can always arise \citep[e.g.,][]{Lund2020} for which one needs to be dark adapted and prepared.  You need to make sure that if you have to apologize to your advisor for losing a star during an observing run, it is because the star is actually gone and not because you cannot see (Steve Wozniak, private communication relayed via Brad Barlow).

\begin{figure*}
    \centering
    \subfloat{
            \includegraphics[width=0.50\textwidth]{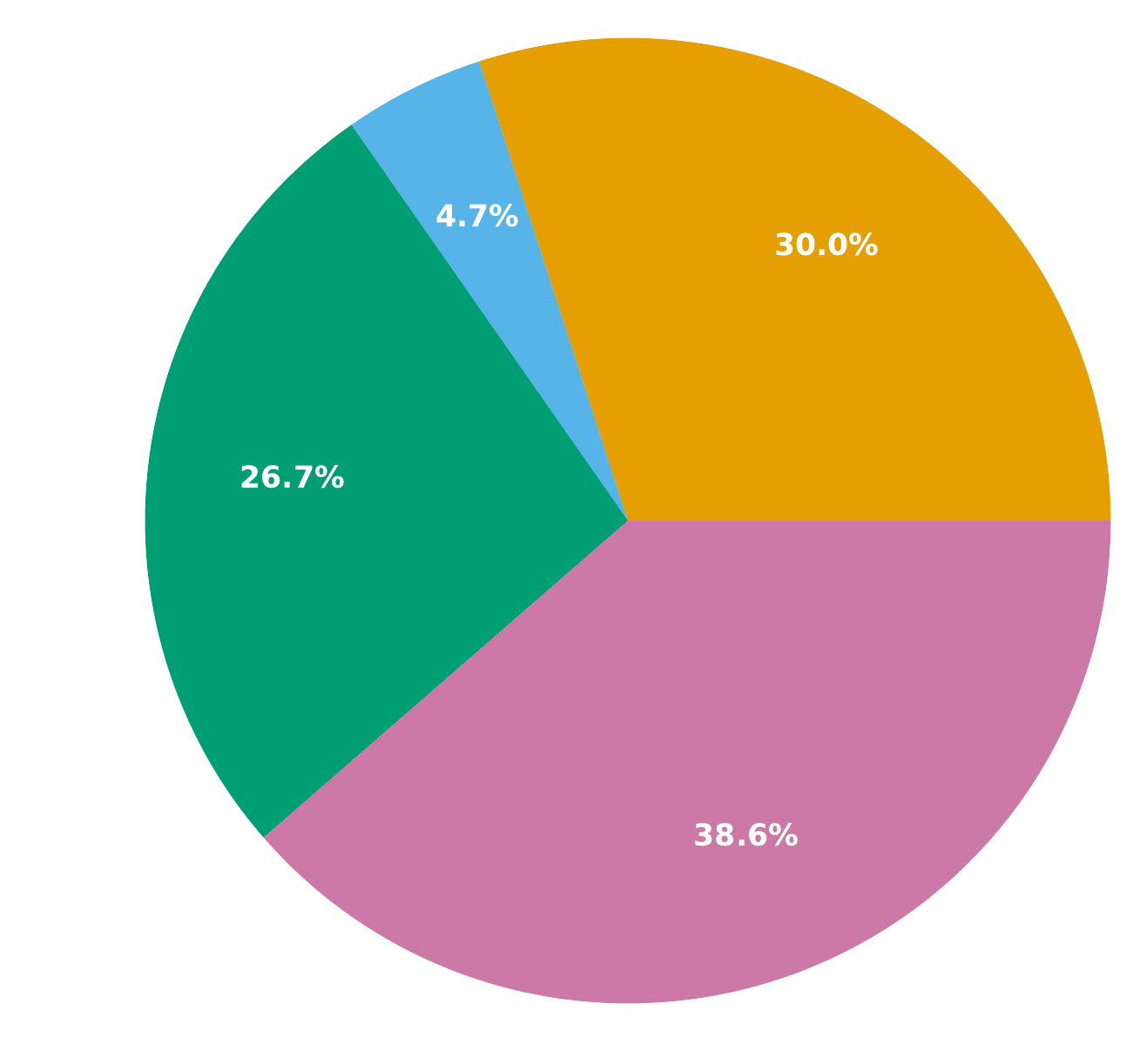}
            }
    \subfloat{
            \includegraphics[width=0.50\textwidth]{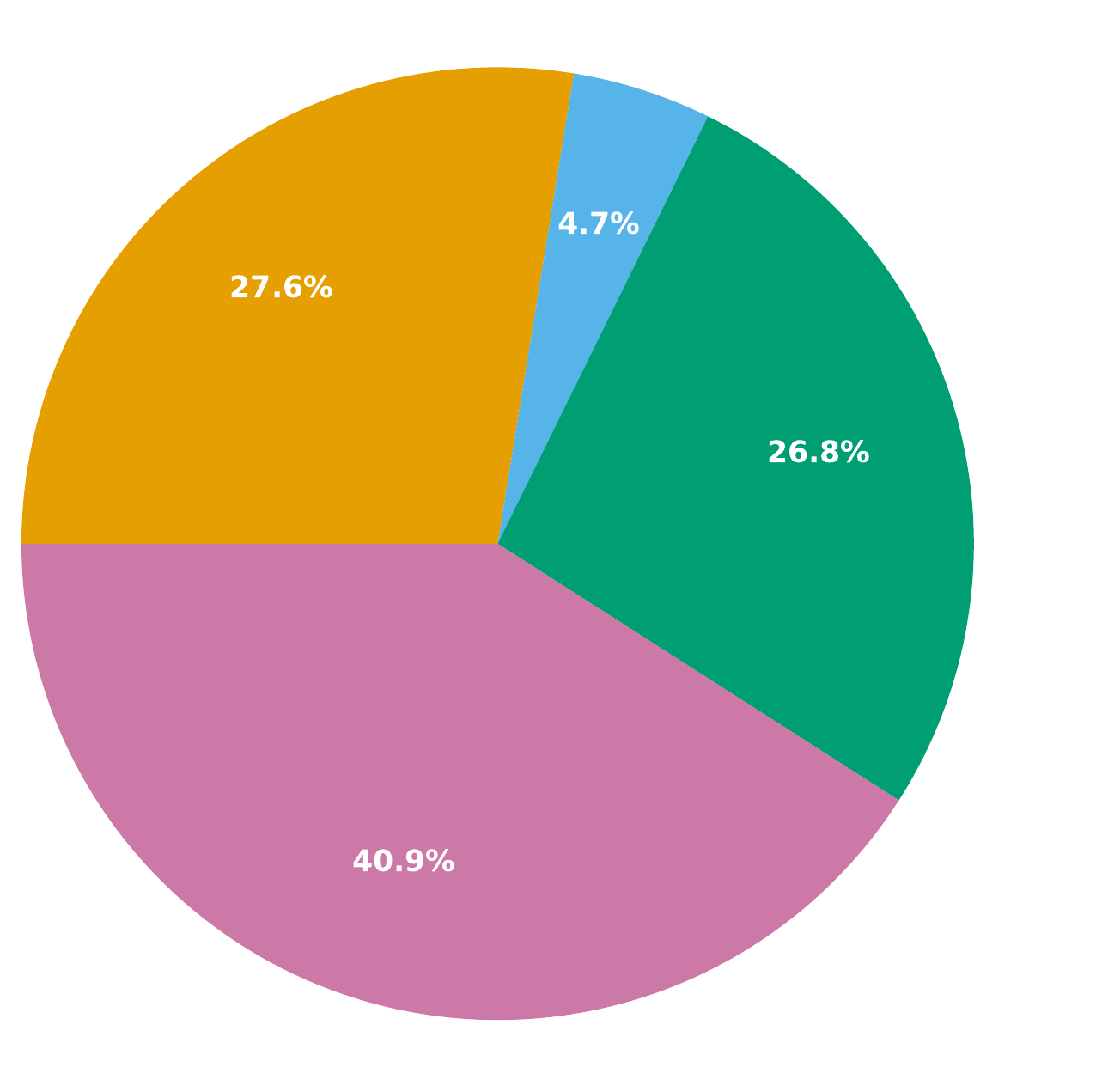}
            } \\
    \caption{Fractional percentages of theme use in iPosters for the two meetings selected.  \textit{Left:} AAS 237, which had a sample of 487 posters.  \textit{Right:} AAS 238, which had a sample of 127 posters.}
    \label{fig:aas_posters}
\end{figure*}

\section{Potentially Predicting Palette Preferences Per Posters} \label{sec:aasposters}
To estimate a rough percentage of users that may find our new DM theme desirable when preparing their manuscripts, we analyzed iPoster design trends from the AAS 237 \& 238 meetings.  This selection allows us not only to determine the total fractional usage of themes in iPosters, but it also allows us to compare the fractional theme usage of winter meeting attendants to that of summer meeting attendants.  Another benefit of these two meetings as a sample is that they were both recent, virtual-only conferences, meaning we expected a larger sample of iPosters than typically at in-person meetings.  We chose to analyze iPosters because we were able to view them, which is a significant benefit in the process of analyzing them.  For this exercise, we define four possible categories of poster design:
\begin{enumerate}
    \item DM -- entirely negative polarity (light text on dark background) poster content
    \item LM -- entirely positive polarity (dark text on light background) poster content
    \item Mixed DM (MDM) -- aspects of both design styles included but predominantly DM poster content
    \item Mixed LM (MLM) -- aspects of both design styles included but predominantly LM poster content
\end{enumerate}
We note that only the text content of the poster was used to determine the design classification as the ability to use a color that provides high contrast to the text boxes or a space-themed image for a poster background adds a non-negligible level of bias towards DM themed titles.  We also note that we discard from our sample posters where the author did not appreciably update the poster design from one of the nine default-template styles as this is another potential source of bias.

\begin{table}
    \centering
    \caption{Breakdown of the themes for iPosters at the two selected AAS meetings.}
    \small
    \begin{tabular}{@{} lrrrrr @{}}
    \toprule
    
    \multicolumn{1}{l}{\normalsize{AAS}} & \multicolumn{1}{c}{\normalsize{DM}} & \multicolumn{1}{c}{\normalsize{LM}} & \multicolumn{1}{c}{\normalsize{MDM}} & \multicolumn{1}{c}{\normalsize{MLM}} & \multicolumn{1}{c}{\normalsize{Discarded}}  \\

    \midrule
    237 & 146 & 130 & 23 & 188 & 275  \\
    238 & 35 & 34 & 6 & 52 & 127  \\
    \hline\hline
    & 181 & 164 & 29 & 240 & 402  \\
    \bottomrule
    
    
    \end{tabular}
    \label{tab:poster_themes}
\end{table}

We visually inspected each poster for both meetings, noting the theme and the topic area of each entry.  In the AAS 237 sample, there were 762 total posters of which 275 were discarded due to a lack of appreciable change from the templates.  For AAS 238, these numbers were 254 and 127, respectively, meaning our overall sample of posters was 614.  A breakdown of the tallied themes for each of the four categories is shown in Table \ref{tab:poster_themes}, and the fractional percentages for the themes at both meetings are shown in Figure \ref{fig:aas_posters}.  We also show a breakdown of DM theme usage by some of the AAS topic areas in Appendix \ref{appendix}.

Perhaps unsurprisingly, LM themes (LM and MLM) were more common than DM themes by about a factor of two overall.  The most common design trend was MLM for both meetings, which can probably be explained mostly in part by the nice contrast of a DM-style banner with a section title above the LM-style text box with content.  What is perhaps most surprising is striking similarity in the fractional percentages of theme choices between the two meetings.  It is possible that there are potential duplicates between the two meetings that would influence our results, but it seems unlikely that there would be enough overlap to cause this similarity.  More data will be needed to compare the fractional theme usage of winter and summer meeting attendants as a result.   Based solely on the total numbers, we would be able to say our new DM theme would be largely appealing to about 35\% of the community; however, due to the large number of MLM posters, there is likely a small subset (we assume $\sim\frac{1}{6}$) of this group that would also find our new theme appealing.  Therefore, we estimate that roughly 42\% of the community would benefit from the addition of a DM journal theme.  

\section{Drawbacks of Using a DM Theme and How to Deal With Them} \label{sec:drawbacks}
Here we describe some of the potential drawbacks that our DM theme could create.  Many of these are not only manageable, but also could not even be issues at all depending on perspective.  We comment on the severity of each disadvantage and how one could mitigate the issues it creates.

\subsection{Printing Papers Out} \label{ssec:printing}
A large disadvantage of a DM theme would be the significant increase in the amount of ink needed to print out a paper.  Ink is very costly, and the length of some manuscripts would strike fear into even the most voluminous of cartridges.  This is quite a severe problem in the context of printing out papers; however, this issue is not an issue should the reader simply not print out the paper and read it online instead.  While this solution does not fit everyone's work style, it is still an option.  Alternatively, authors and journals could provide both a LM and DM version of the paper when publishing, allowing user preference to resolve this issue naturally.  Obviously, this also comes with the added time associated with preparing plots in both styles; however, this once again highlights the strength of the CBF colors described in \textsection\ref{ssec:links} and \textsection\ref{sssec:cbfcolors}.

\subsection{Legibility and Reading Comprehension} \label{ssec:reading}
\subsubsection{The ``Positive Polarity Advantage''} \label{sssec:ppadvantage}
A fairly well researched advantage that LM themes have is the legibility and comprehension of reading dark text on a light background \citep[see introduction of][and references therein]{DOBRES2017}, especially with small text.  When reading text in LM, the reader's pupil contracts more making the text easier to read and this leads to a better comprehension.  This has been dubbed in the literature the ``positive polarity advantage'' due to the benefits that LM has on long reading sessions; however, \citet{DOBRES2017} found that in many different ambient light conditions, the increase in legibility and comprehension was not significantly higher.  They thus suggested that the positive polarity advantage is likely more of a slight ``negative polarity disadvantage.''  Based on the results of this study, we do not foresee any issues with our DM theme in most lighting conditions.  We also do not believe this will be a severe disadvantage to using DM for this reason.

\subsubsection{Low-light Conditions} \label{sssec:lowlight}
The one case of ambient lighting that \citet{DOBRES2017} did find a significant disparity between LM and DM reading was in settings with dark illumination.  Here dark text on a light background was much easier to read and comprehend, which seems to be expected for an extended reading session in low-light conditions.  Reading for prolonged periods of time in low-light conditions in general is tiring, too.  While there is not truly a good way to mitigate this mildly severe issue, we have touched on reasons one might still wish to use a DM theme in low-light conditions in \textsection\ref{ssec:rhodopsin}.  As stated in \textsection\ref{ssec:printing}, having both LM and DM options for submissions would also be useful in resolving this issue.

\subsection{People Might Just Hate Our Proposed Theme} \label{ssec:hatethetheme}
We recognize that the theme we have developed may not be a long-term solution, and we also realize people may severely dislike the specific design choices made here.  Many of these criticisms could be entirely valid; however, we remind the reader that the main premise of this text is that a DM theme should be introduced to several journals.  That does not mean it has to be the theme we present here, meaning there is always room for improvement in the design.  Our theme is simply one of what could be many potential options.  This particular issue might be severe for some while for others it is not, which is another reason for offering both DM and LM.

\vspace{1mm}
\section{Summary} \label{sec:summary}
We have presented the case for the addition of ``dark mode'' to journals, in particular arXiv pre-prints and those under the AAS Journal umbrella.  Our modified class file makes generating a DM-themed paper simple, and our adopted color schemes for the background, text, and links make the papers look phenomenal.  There are many benefits to using our DM theme, including increased battery life, reduced screen burn-in, increased efficiency when figure hopping, the useful implementation of a CBF color palette, and saving rhodopsin during observing runs.  Trends in AAS iPoster design from recent AAS meetings suggest that a significant portion of the astronomical community would benefit from or at least have interest in this new theme.  As with all good things, our DM theme does have some drawbacks; however, many of these are manageable by the user.  The addition of a DM theme -- not necessarily even the one presented here -- has the potential to make us all better astronomers for these reasons, so we urge those in the community to consider this possibility.

\section*{Acknowledgments}\label{acknowledgments}
We thank Mark Siebert, Connor McClellan, and Xinlun Cheng for their insights and suggestions that helped improve the content of this paper.  We thank all the Astro Pups and Cats we have interacted with during the time this manuscript was being prepared.  We acknowledge the many conversations that took place in UVA Astronomy office 263 on the subjects discussed here that led to the creation of this manuscript.  We also acknowledge the support and helpful comments from the Barlow Research Group Alumni Virtual Happy Hour.  We lightly curse \texttt{xgterm} for not providing a functional DM alternative or really any customization/usability changes for use with IRAF.  We thank the AAS iPoster session URLs for being predictable yet still functional to allow data to be collected for this work -- they did not have to do that but did anyway.  We thank \textit{Super Auto Pets} for providing the unmatched fun during breaks we needed in 2022.  We thank Graco\textsuperscript{\textregistered} for the Simple Sway\textsuperscript{\texttrademark} Swing and the added productivity it brings us.  Finally, we thank Walt Disney World for providing not only a magical experience each and every time we go, but also for our never-ending references to ``I'm going to Disney World!'' following the publication of papers. 

%

\facilities{Dark Room at FMO:40in, UVa Astronomy Building:Office263, The Living Room}


\software{GIMP \citep{gimp}, \texttt{NumPy} \citep{harris2020array}, \texttt{Matplotlib} \citep{Hunter2007}
         }



\appendix
\section{iPoster DM Theme Usage by AAS Topic Area} \label{appendix}
Using the combined data from both AAS meetings, we plot the total number of DM theme posters by common, named subject areas at AAS meetings in Figure \ref{fig:topics}.  This large plot showcases many of the named colors in \texttt{Matplotlib} that cannot be used or do not look particularly good in a traditional LM document.  We have further breakdowns of the DM, LM, MDM, MLM, and discarded tallies by subject area and AAS meeting that can be made available upon request; however, those are outside the purview of this paper at this time.
\begin{figure}
    \centering
    \includegraphics[width=\textwidth]{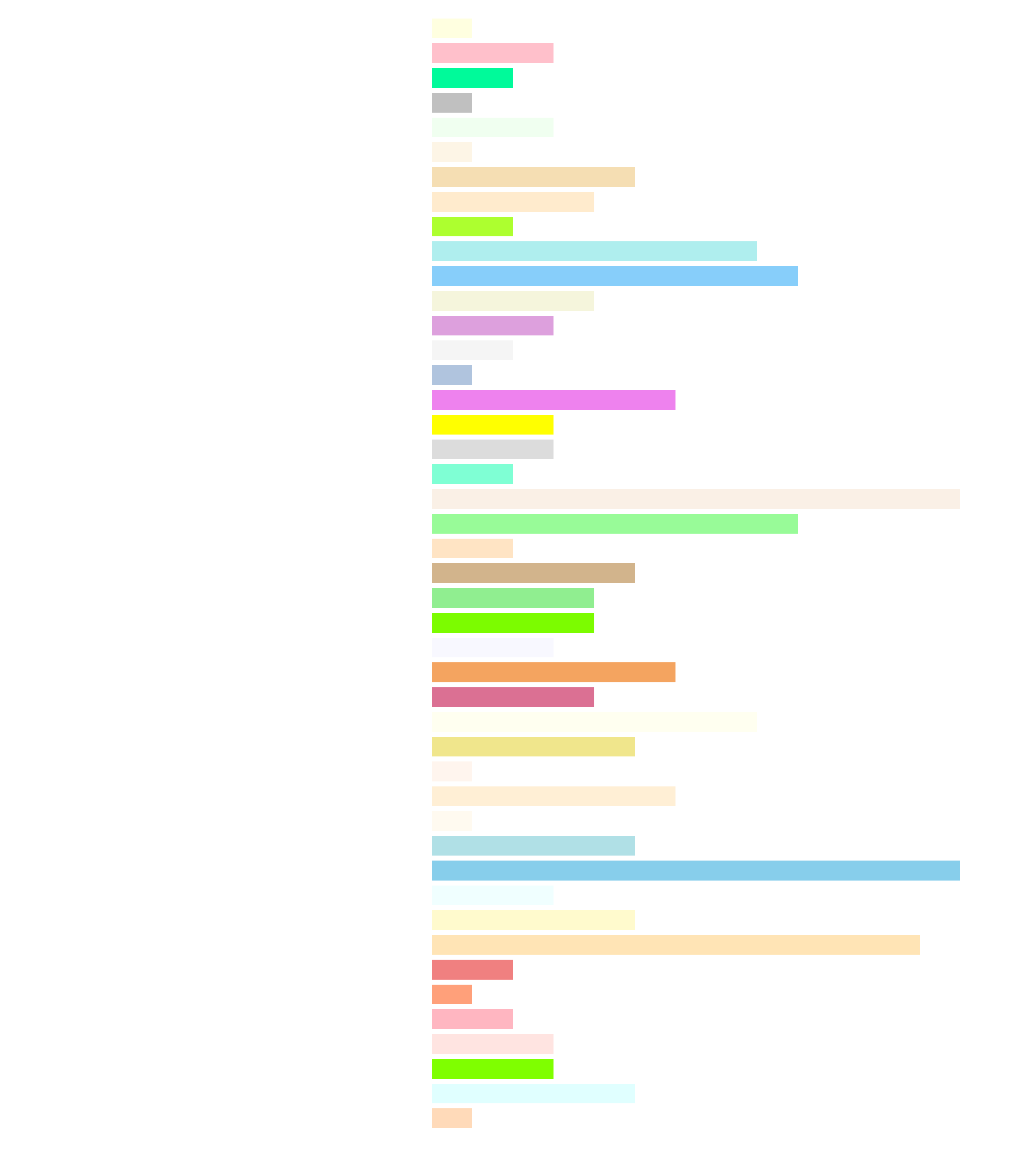}
    \caption{A tally of the number of iPosters from various subject areas at AAS meetings 237 \& 238 that used a DM theme.  Many of the newly justified colors for plotting are used in this figure to showcase how they work in a DM document.}
    \label{fig:topics}
\end{figure}


\bibliography{sample631}{}
\bibliographystyle{aa_url_dm}



\end{document}